\documentclass[11pt,twoside]{article}

\usepackage{asp2004}
\usepackage{epsf}
\usepackage{psfig}
\usepackage{lscape}
\usepackage{graphicx}
\usepackage{natbib}

\begin{document}
\title{Infrared observations of V838~Mon}
\author{M. T. Rushton}  
\affil{Astrophysics Group, Keele University, Keele, Staffordshire, ST5 5BG, UK}   
\author{T. R. Geballe}   
\affil{Gemini Observatory, 670 N. A'ohoku Place, Hilo, HI\,96720, USA}
\author{A. Evans, J. Th. van Loon, B. Smalley} 
\affil{Astrophysics Group, Keele University, Keele, Staffordshire, ST5 5BG, UK}    
\author{S. P. S. Eyres} 
\affil{Centre for Astrophysics, University of Central Lancashire,
      Preston, PR1 2HE, UK}   

\begin{abstract} 
We describe the results of fitting simple spherically symmetric models to
the first overtone CO and AlO A--X (2--0) bands in V838~Mon. We find that
the temperature and column of both CO and AlO systematically decline over
the period 2002 October -- 2005 February and that an additional, hotter
and denser, component is present from 2005 January. We also describe the
results of an observation of the star and its light echo at 850\micron.
We do not detect the `infrared' echo at these wavelength, and place an
upper limit of $3\times10^7$~cm$^{-2}$ on the column of grains.
\end{abstract}

\newcommand{\vunit}{\mbox{~km~s$^{-1}$}}

\section{Introduction}

We began a programme of regular infrared (1--5\micron) spectroscopic
monitoring of V838~Mon soon after the 2002 eruption was reported by
\cite{brown};
the results are described in \cite{evans,rushton-mega,rushton-sio,phd}.
All of the data reported here cover the first overtone CO bands
(bandhead at 2.29\micron)
and the A--X (2--0) AlO bands at 1.67\micron. In this contribution we
discuss our attempts to model the complex evolution of these bands, as
revealed by our low resolution obervations of CO and AlO.

Details of the infrared data we obtained at high resolution, and what
the data tell us about the complex velocity systems in the environment
of V838~Mon, is discussed by Geballe elsewhere in these proceedings. 
 
We also describe here the results of an observation at 850\micron\ with
the James Clerk Maxwell Telescope (JCMT).

\section{Modelling CO and AlO}

\begin{figure}[!ht]
\plotfiddle{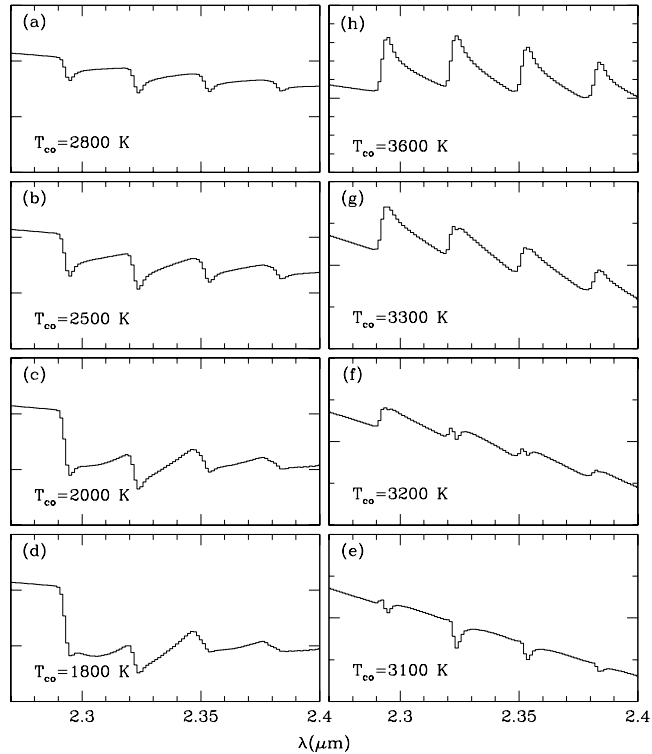}{95mm}{0}{50}{50}{-150}{-70}
\caption{Model spectra around the first overtone $^{12}$CO bands.
CO column $= 10^{21}$~cm$^{-2}$ and excitation temperature $T_{\rm CO}$
as shown \cite{phd,rushton-co}.}
\label{CO}
\end{figure}

We consider the star to be at the centre of one or two spherically
symmetric, expanding shell(s). As noted by Geballe (this volume)
it is more likely that three, or even four, shells are present;
however at the low resolution of the spectra one or two shells
gives a good representation of the data. We have taken the first
overtone CO linelist (in excess of 40000 lines) from \cite{goorvitch}
and AlO data from various sources (see \cite{phd} for details). 

We fit the models to the data using an AMOEBA algorithm, based on
downhill simplex \cite{press}. We optimized for the excitation
temperature and column of CO, and similarly for AlO, and for the
temperature of the background (blackbody) continuum.
For the present we assume LTE. We found that the
temperature of the background varied between $\sim4000$~K and
$\sim2000$~K. Initially we also optimized for the $^{12}$C/$^{13}$C
ratio but this was not well constrained, with a lower limit of 16;
this is consistent with the apparent absence of $^{13}$CO bands
described by Geballe (this volume).

Optimization (particularly for AlO) is not helped by the presence of
many other species, particularly H$_2$O, which are not included in
our analysis at this stage.

\section{Results}
\subsection{Models}

Examples of model spectra around the first overtone CO bands are 
shown in Fig.~\ref{CO}.
The model spectra qualitatively cover the range of first overtone
CO profiles observed in V838~Mon by our programme and by
\cite{banerjee}. At 2800~K the CO spectra are qualitatively similar
to those seen in late giants. At low enough CO excitation
temperature ($T_{\rm CO}$) the optical depth in the CO lines is
$\sim$ constant across the band and the bands lose their
characteristic asymmetric shape. At $T_{\rm CO}=2000$~K the model
bands resemble data obtained in our programme, those at 1800~K
resemble data reported by \cite{banerjee}. At high excitation
temperatures the CO begins to appear in emission. In particular
the 1800~K model is strikingly similar to the CO profiles for
2002 October and, while the models with $T_{\rm CO}>3000$~K
produce emission, the profiles are not flat as they are in the
2002 March data.

\subsection{Fits}

\begin{figure}[!ht]
\plottwo{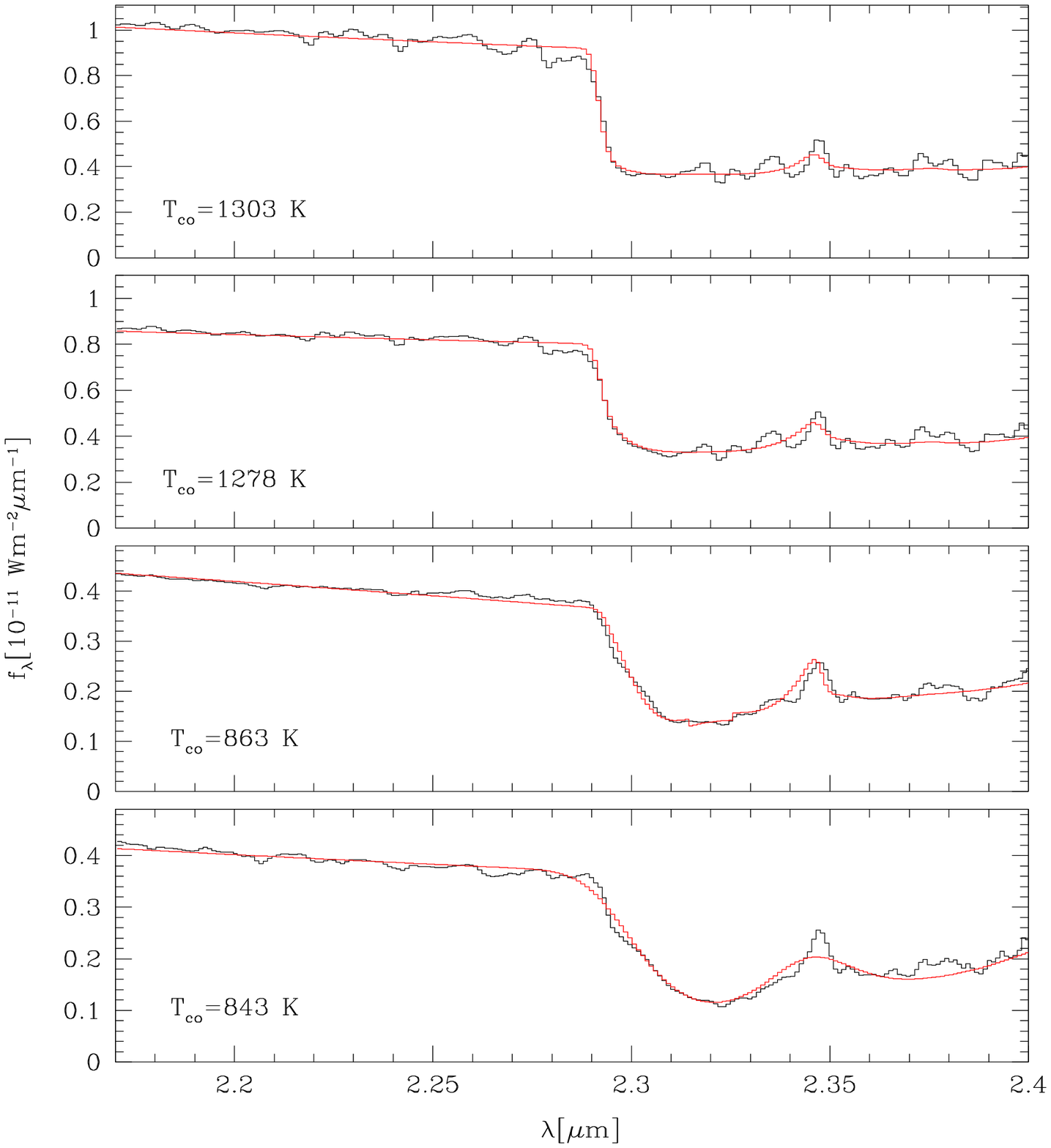}{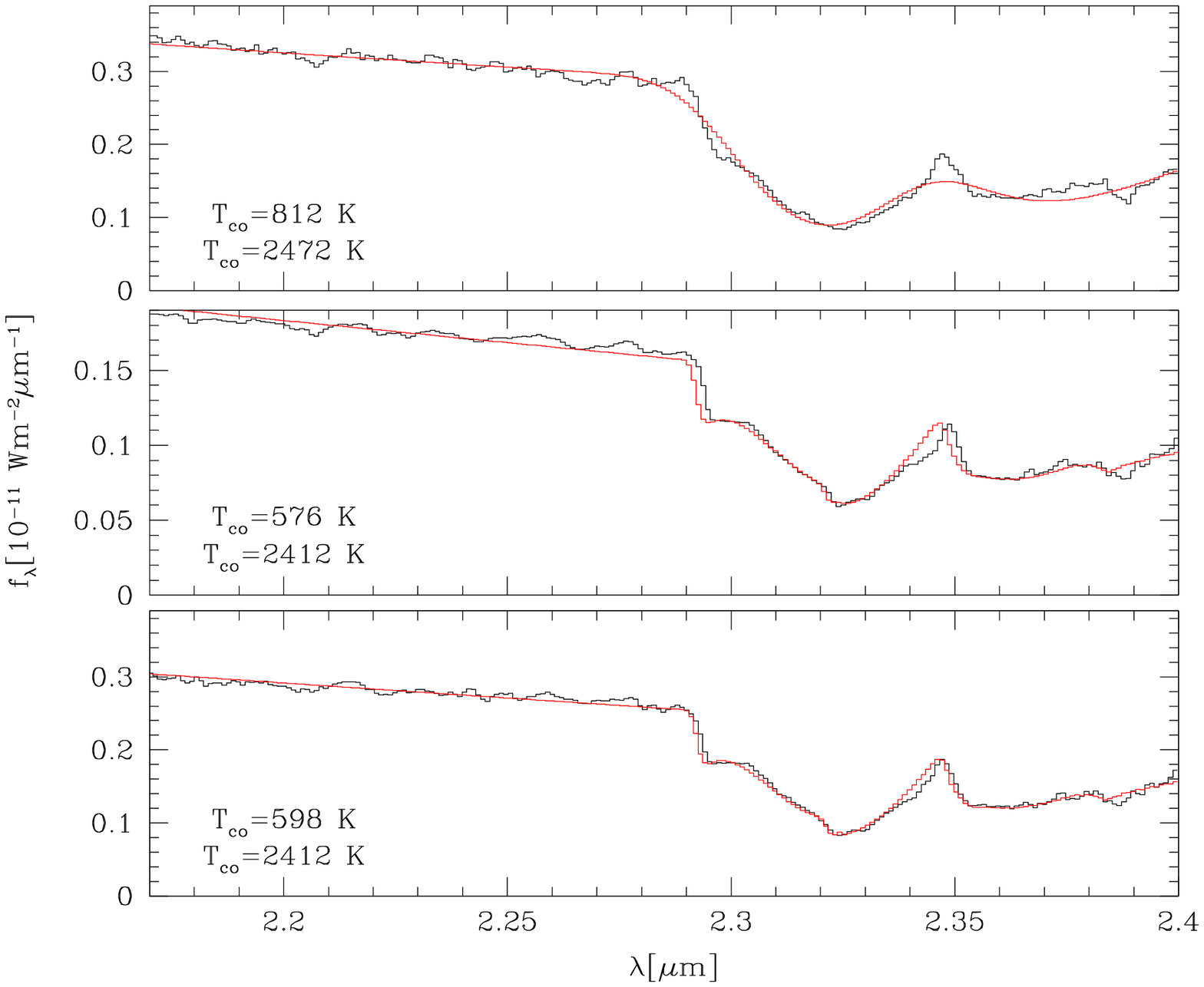}
\caption{Model fits to the CO data for 2002 October -- 2005 February.}
\label{CO1}
\end{figure}

\begin{figure}[!ht]
\plottwo{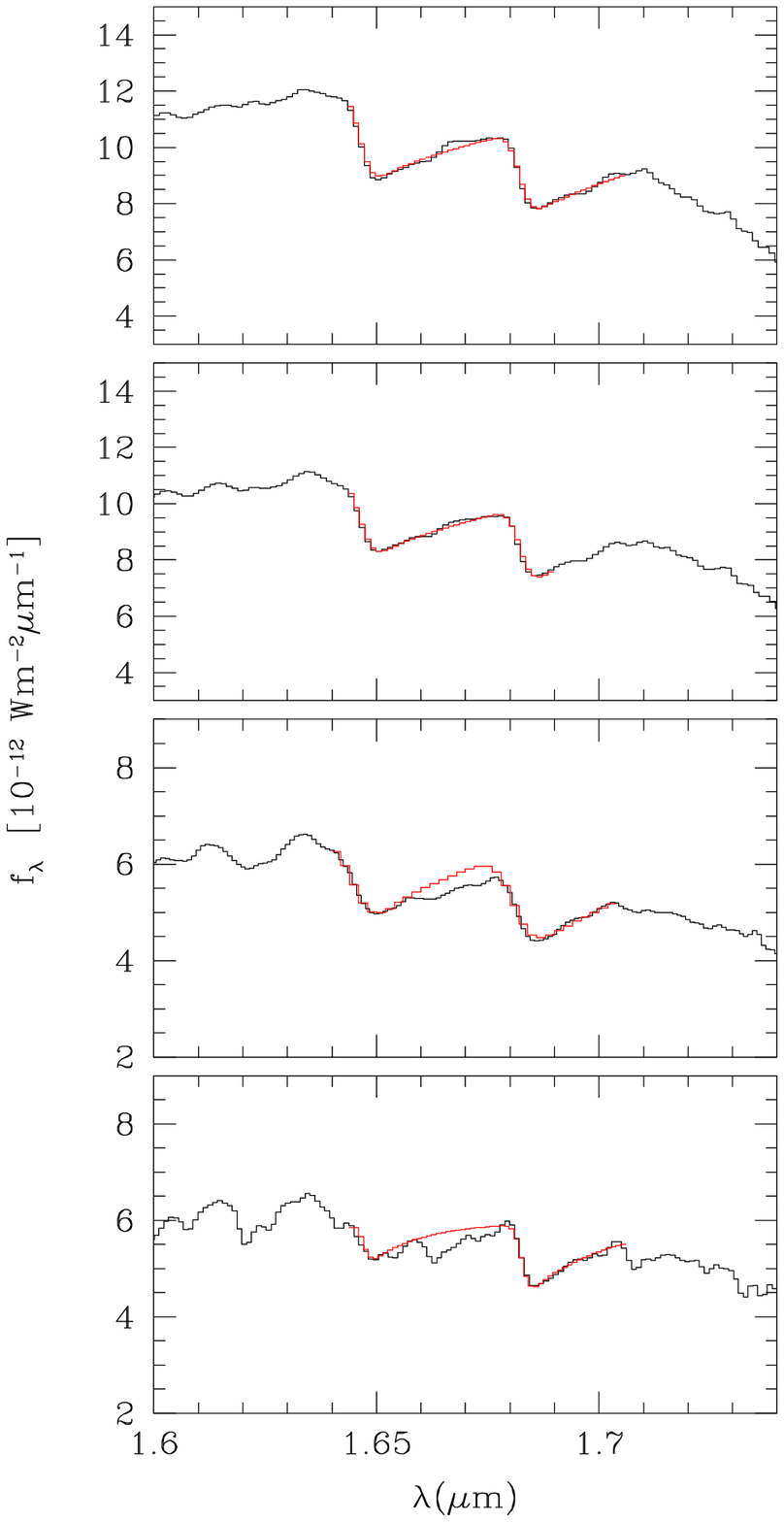}{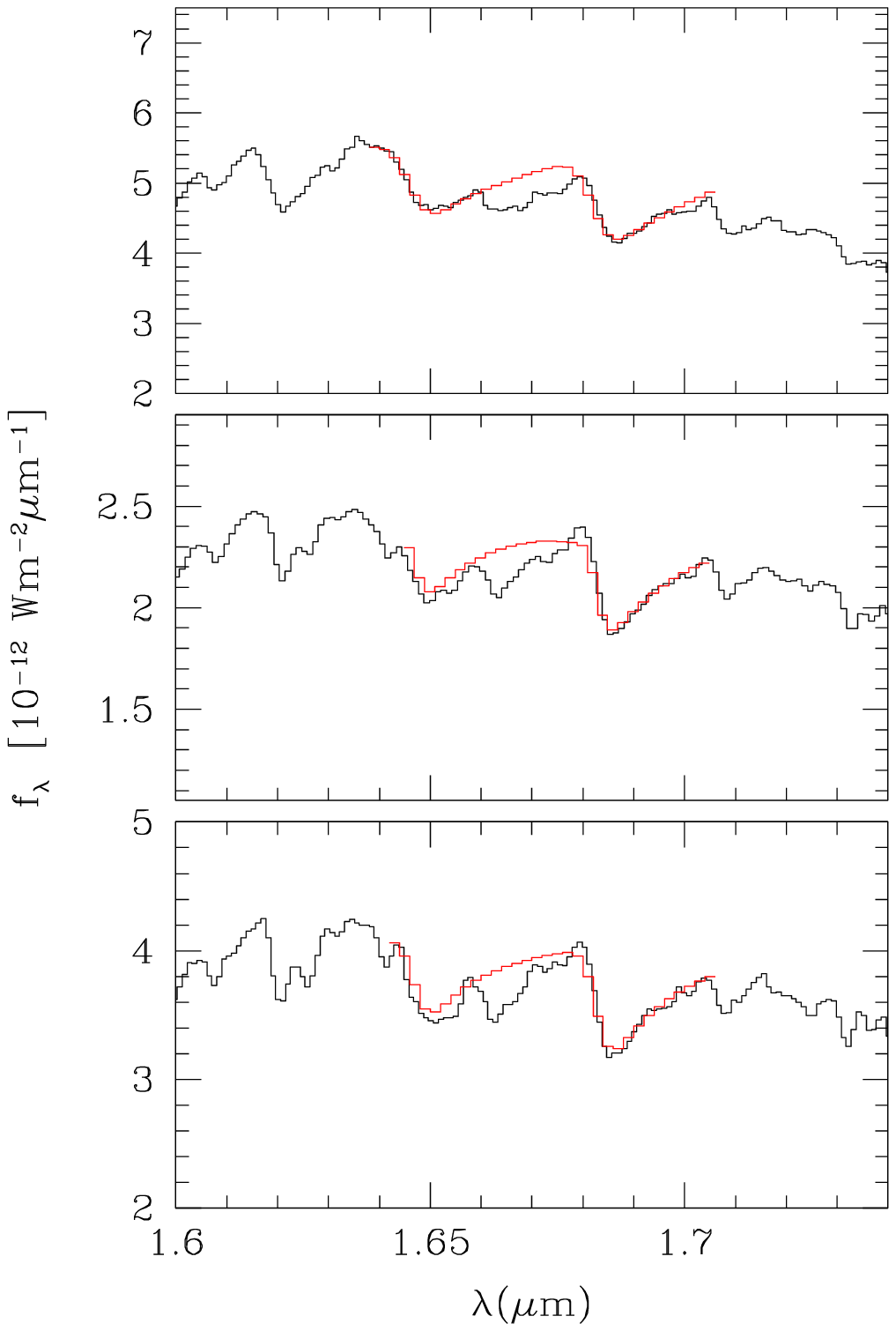}
\caption{Model fits to the AlO data for 2002 October -- 2005 February.}
\label{alo}
\end{figure}

\begin{figure}[!ht]
\plotfiddle{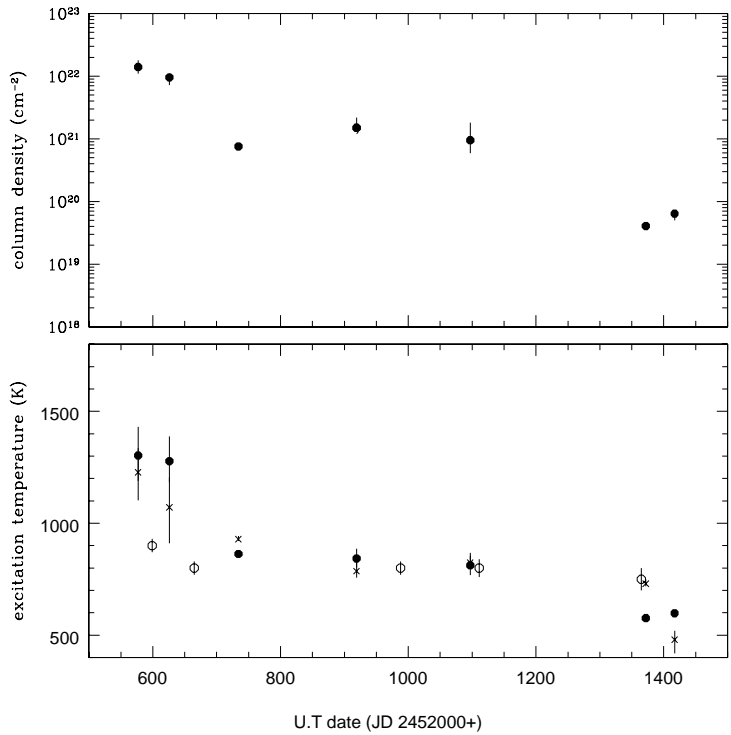}{90mm}{0}{50}{50}{-150}{-80}
\caption{Decline of CO column (top) and excitation temperature
for CO, AlO and H$_2$O (bottom) with time;
$\bullet=$CO, $\circ-$H$_2$O \cite{barber}, $\times=$AlO.}
\label{CO2}
\end{figure}

The fits to the first overtone CO bands in V838~Mon are shown in
Fig.~\ref{CO1}; the fits to the AlO bands are shown in Fig.~\ref{alo}.
We find that the excitation temperature and column of CO decline with
time, as shown in Fig.~\ref{CO2}, presumably the result of the
detachment, followed by cooling and dispersal, of material ejected
in the 2002 event. The run of CO excitation temperature and column
is broadly in line with that found by \cite{lynch}; furthermore the
decline in the excitation temperature, for both CO and AlO, is consistent
with that found by \cite{barber} for H$_2$O.

We note that there is clear evidence at the band head for an
additional, hotter component in the later data; this may be the
photosphere reappearing through the ejected material.

While our simple model provides a good qualitative account of the
behaviour of the CO and AlO we consider that a more complete
description will likely require more shells than we have considered here
(cf. Geballe, this volume) and non-LTE treatment.

\section{Observation at 850\micron}

There are benefits in observing the infrared equivalent
of light echoes, for example the fact that detailed knowledge of
the grain scattering phase function is not required for
interpretation. In this case we see the `infrared echo',
commonly observed (as indeed are light echoes) when supernovae
erupt \cite{irecho}. In the infrared case we see the results
of grain heating and re-radiation as the paraboloid of
constant light travel time (see Sugerman, this volume)
passes through the dust. The thermal inertia of grains
is such that they react essentially instantaneously to changes in
the ambient radiation field.

\cite{banerjee} has presented Spitzer MIPS observation of the
V838~Mon infrared echo; see also Ashok (this volume). They
find that the emission at 24, 70 and 160\micron\ is spatially
correlated with the HST light echo; they also find emission
associated with the dust located close to the star itself.
They estimate that the mass of dust in the light echo material
is $\sim$~a few solar masses, and they conclude that a significant
fraction of the emitting dust is likely interstellar in origin.

\begin{figure}[!ht]
\plotfiddle{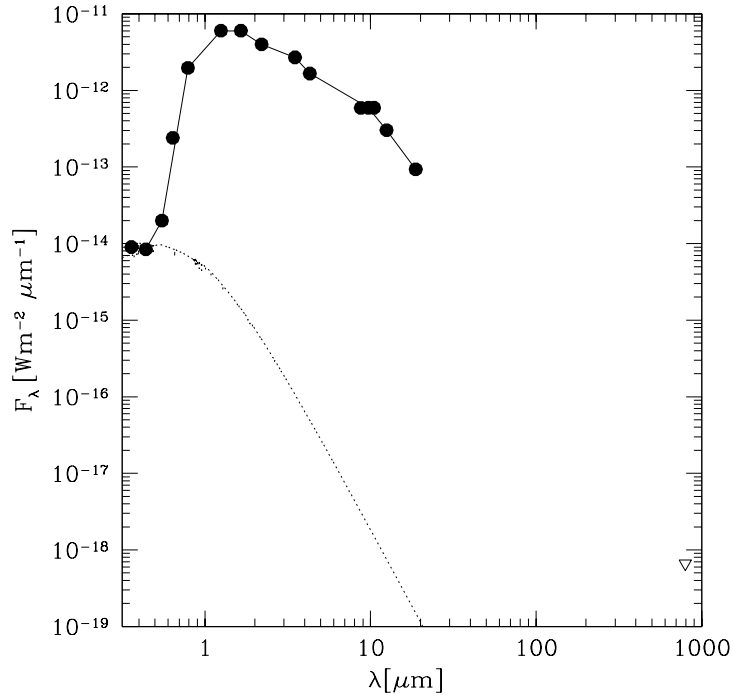}{90mm}{0}{48}{48}{-150}{-80}
\caption{The spectral energy distribution of V838~Mon from late 2003.
Filled circles, data from \cite{crause5, wagner,lynch,tapia}.
The $U\!BV\!RI$ data are from 2003 September, while the remainder are
from 2003 October. Triangle: our upper limit at 850\micron.
The dotted line is a B3V star, reddened by $E(B-V)=0.85$ and
scaled to match the $U$ and $B$ bands.}
\label{sed}
\end{figure}

We have used the Sub-millimetre Common User Bolometer Array (SCUBA)
on the JCMT to make a $2'\times2'$ 850\micron\ jiggle map of the
light echo over the period 2003 October --  2004 January
(most of the observations being in 2004 January), when the
radius of the light echo was $\simeq53''$ \cite{crause}.
We get a $3\sigma$ upper limit of 160mJy from the star itself;
the SED for late 2003 is shown in Fig.~\ref{sed}.

We find a $3\sigma$ upper limit at 850\micron\ of
$2\times10^4$~Jy~sr$^{-1}$ from the region containing the light echo
material. In order to interpret this result we have estimated the
likely intensity of the 850\micron\  emission from the light echo
dust by making an analytic fit to the time-dependence of the
bolometric luminosity $L_*(t)$ and $T_*(t)$ from \cite{tylenda}
and integrating through a plane thin dusty inclined slab \cite{crause5}.
We take optical constants for silicates from \cite{draine}.

We find an upper limit on the dust column in the echo material
$3\times10^7$~cm$^{-2}$. This column gives an upper limit of
$\la 0.1$M$_{\odot}$, within $50\arcsec$, for 0.1\micron\ silicate
grains and for 5--9~kpc distance; this is clearly significantly
below the mass deduced by \cite{spitzer}.

\section{Conclusions}

We have fitted a spherically symmetric, expanding shell of CO and AlO to
the near-infrared spectra of V838~Mon. We find a declining column of CO
and AlO, and a declining $T_{\rm ex}$ for both species. However there is
an additional hot CO component from 2005 Jan.

We find an upper limit of $3\times10^7$~cm$^{-2}$, and corresponding
upper mass limit of $0.1$~M$_\odot$, for the dust in light echo
material; this is significantly lower than that implied by the
Spitzer observations.

Further details of this work will be published elsewhere.

\acknowledgements MTR was supported by the UK Particle Physics
\& Astronomy Research Council (PPARC)
and is now supported by the University of Central Lancashire.
TRG is supported by the Gemini Observatory, which is operated by the Association
of Universities for Research in Astronomy, Inc., on behalf of the international
Gemini partnership of Argentina, Australia, Brazil, Canada, Chile, the United
Kingdom, and the United States of America.

%%% THE BIBLIOGRAPHY
%%%
%%% CONSULT SECTION 4 OF "INSTRUCTIONS FOR AUTHORS" FOR HOW TO USE NATBIB.
%%% PLEASE USE THE "THEBIBLIOGRAPY" ENVIRONMENT
%%%

% edit and fill the lines below with the questions and answers, or 
% comment all lines except \end{document}

\end{document}